\documentstyle[12pt]{article}

\catcode`\@=11
\long\def\@makefntext#1{ 
\protect\noindent \hbox to 3.2pt {\hskip-.9pt
$^{{\ninerm\@thefnmark}}$\hfil}#1\hfill} 

 \def\@makefnmark{\hbox to 0pt{$^{\@thefnmark}$\hss}}  

\def\ps@myheadings{\let\@mkboth\@gobbletwo
\def\@oddhead{\hbox{} 
\rightmark\hfil\ninerm\thepage}
\def\@oddfoot{}\def\@evenhead{\ninerm\thepage\hfil 
\leftmark\hbox{}}\def\@evenfoot{}
\def\sectionmark##1{}\def\subsectionmark##1{}}

\newcounter{sectionc}\newcounter{subsectionc}\newcounter{subsubsectionc}
\renewcommand{\section}[1] {\vspace{0.6cm}\addtocounter{sectionc}{1}
\setcounter{subsectionc}{0}\setcounter{subsubsectionc}{0}\noindent
        {\bf\thesectionc. #1}\par\vspace{0.4cm}}
\renewcommand{\subsection}[1] {\vspace{0.6cm}\addtocounter{subsectionc}{1}
        \setcounter{subsubsectionc}{0}\noindent
        {\it\thesectionc.\thesubsectionc. #1}\par\vspace{0.4cm}}
\renewcommand{\subsubsection}[1]
{\vspace{0.6cm}\addtocounter{subsubsectionc}{1}
        \noindent {\rm\thesectionc.\thesubsectionc.\thesubsubsectionc.
        #1}\par\vspace{0.4cm}}

\newcounter{appendixc}
\newcounter{subappendixc}[appendixc]
\newcounter{subsubappendixc}[subappendixc]

\renewcommand{\appendix}[1] {\vspace{0.6cm}
        \refstepcounter{appendixc}
        \setcounter{figure}{0}
        \setcounter{table}{0}
        \setcounter{equation}{0}
        \renewcommand{\thefigure}{\Alph{appendixc}.\arabic{figure}}
        \renewcommand{\thetable}{\Alph{appendixc}.\arabic{table}}
        \renewcommand{\theappendixc}{\Alph{appendixc}}
        \renewcommand{\theequation}{\Alph{appendixc}.\arabic{equation}}
        \noindent{\bf Appendix \theappendixc #1}\par\vspace{0.4cm}}

\def\abstracts#1{{
        \centering{\begin{minipage}{30pc}\tenrm\baselineskip=12pt\noindent
        \centerline{\tenrm ABSTRACT}\vspace{0.3cm}
        \parindent=0pt #1
        \end{minipage} }\par}}

\renewenvironment{thebibliography}[1]
        {\begin{list}{\arabic{enumi}.}
        {\usecounter{enumi}\setlength{\parsep}{0pt}
\setlength{\leftmargin 1.25cm}{\rightmargin 0pt}
         \setlength{\itemsep}{0pt} \settowidth
        {\labelwidth}{#1.}\sloppy}}{\end{list}}

\newcounter{itemlistc}
\newcounter{romanlistc}
\newcounter{alphlistc}
\newcounter{arabiclistc}

\newcommand{\fcaption}[1]{
        \refstepcounter{figure}
        \setbox\@tempboxa = \hbox{\tenrm Fig.~\thefigure. #1}
        \ifdim \wd\@tempboxa > 6in
           {\begin{center}
        \parbox{6in}{\tenrm\baselineskip=12pt Fig.~\thefigure. #1 }
            \end{center}}
        \else
             {\begin{center}
             {\tenrm Fig.~\thefigure. #1}
              \end{center}}
        \fi}

\newcommand{\tcaption}[1]{
        \refstepcounter{table}
        \setbox\@tempboxa = \hbox{\tenrm Table~\thetable. #1}
        \ifdim \wd\@tempboxa > 6in
           {\begin{center}
        \parbox{6in}{\tenrm\baselineskip=12pt Table~\thetable. #1 }
            \end{center}}
        \else
             {\begin{center}
             {\tenrm Table~\thetable. #1}
              \end{center}}
        \fi}

\def\@citex[#1]#2{\if@filesw\immediate\write\@auxout
        {\string\citation{#2}}\fi
\def\@citea{}\@cite{\@for\@citeb:=#2\do
        {\@citea\def\@citea{,}\@ifundefined
        {b@\@citeb}{{\bf ?}\@warning
        {Citation `\@citeb' on page \thepage \space undefined}}
        {\csname b@\@citeb\endcsname}}}{#1}}

\newif\if@cghi
\def\cite{\@cghitrue\@ifnextchar [{\@tempswatrue
        \@citex}{\@tempswafalse\@citex[]}}
\def\citelow{\@cghifalse\@ifnextchar [{\@tempswatrue
        \@citex}{\@tempswafalse\@citex[]}}
\def\@cite#1#2{{$\null^{#1}$\if@tempswa\typeout
        {IJCGA warning: optional citation argument
        ignored: `#2'} \fi}}

\def\fnt#1#2{\footnotetext{\kern-.3em
        {$^{\mbox{\sevenrm #1}}$}{#2}}}

 1
\font\twelverm=cmr10 scaled\magstep 1
 1

\font\tenbf=cmbx10
\font\tenrm=cmr10
\font\tenit=cmti10

\font\ninerm=cmr9

\begin{document}
\input epsf
\centerline{\tenbf Hard thermal loop resummation techniques}
\baselineskip=18pt
\centerline{\tenbf in hot gauge theories}
\vspace{0.5cm}
\centerline{\tenrm Randy Kobes}
\baselineskip=17pt
\centerline
{\tenit Physics Department and Winnipeg Institute for Theoretical Physics}
\baselineskip=12pt
\centerline{\tenit University of Winnipeg}
\baselineskip=12pt
\centerline{\tenit Winnipeg, Manitoba\ R3B 2E9\ Canada}
\vspace{0.9cm}
\abstracts{A review is given of the hard thermal loop resummation
methods initiated by Braaten, Pisarski, and others.
We describe some successes of these techniques as well as
instances where modifications are necessary. Some possible directions
where these modifications may lead are also discussed.}
\vfil
\twelverm   
\baselineskip=14pt
\par\section{Introduction}
\begin{figure}[hb]
\begin{center}
\leavevmode
\epsfxsize=4.5 in
\epsfbox{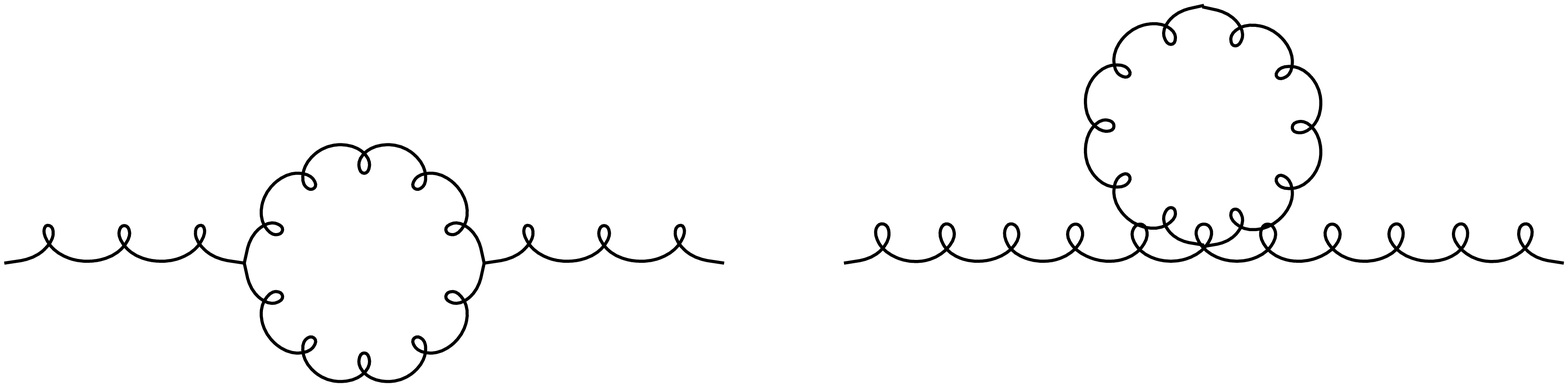}
\end{center}
\fcaption{One--loop gluon self--energy}
\label{oneloop}\end{figure}

The infrared behaviour of gauge theories at high temperature
is generally more pronounced than at zero temperature
and leads to a number of interesting problems.\cite{michel}
One of the outstanding paradoxes at the time of the first workshop
in this series in Cleveland in 1988 was the gauge dependence of the
one--loop gluon damping constant.\cite{r1} The problem at the time was the
following. If one considers the response of a field to a small
external perturbation $J(t)$,
\begin{equation}
  \delta\phi(t)\sim\int dt'\,D(t-t')J(t')\sim
  \int dt'\int dk_0e^{-ik_0(t-t')}D(k_0)J(t'),
\label{response}\end{equation}
then the poles of the (retarded) propagator $D(k)$,
\begin{equation}
  D(k)\sim \frac{i}{k_0-\Sigma(k)}\sim
  \frac{i}{k_0-\omega(k)+i\gamma(k)},
\end{equation}
will determine a characteristic frequency $\omega$ and
damping rate $\gamma$ of the oscillations. For $QCD$ in the
high temperature, long wavelength limit one can calculate
these quantities at the one--loop level by considering the
graphs in Fig.~\ref{oneloop}, along with Faddeev--Popov ghost
terms as required;
\par\noindent
the frequency
\begin{equation}
  \omega(k)\sim {\rm Re}\, \Sigma(k_0=\omega)
\end{equation}
turns out to be of order $gT$ and is gauge independent,
but the damping rate
\begin{equation}
  \gamma\sim{\rm Im}\, \Sigma(k_0=\omega)
\end{equation}
is of order $g^2T$ but gauge dependent.\cite{r2}${}^{-}$\cite{r7}
 There was even some
controversy over the sign of $\gamma$, which from
Eq.(\ref{response}) would determine if the oscillations
were damped or anti--damped in time. These results for $\gamma$
were puzzling because,
as well as being a supposedly physical quantity,
there exist formal arguments that the
poles of a propagator are generally gauge independent, even
if the propagator itself is not.\cite{thooft}${}^-$\cite{proof}
\par
\begin{figure}[hb]
\begin{center}
\leavevmode
\epsfxsize=6 in
\epsfbox{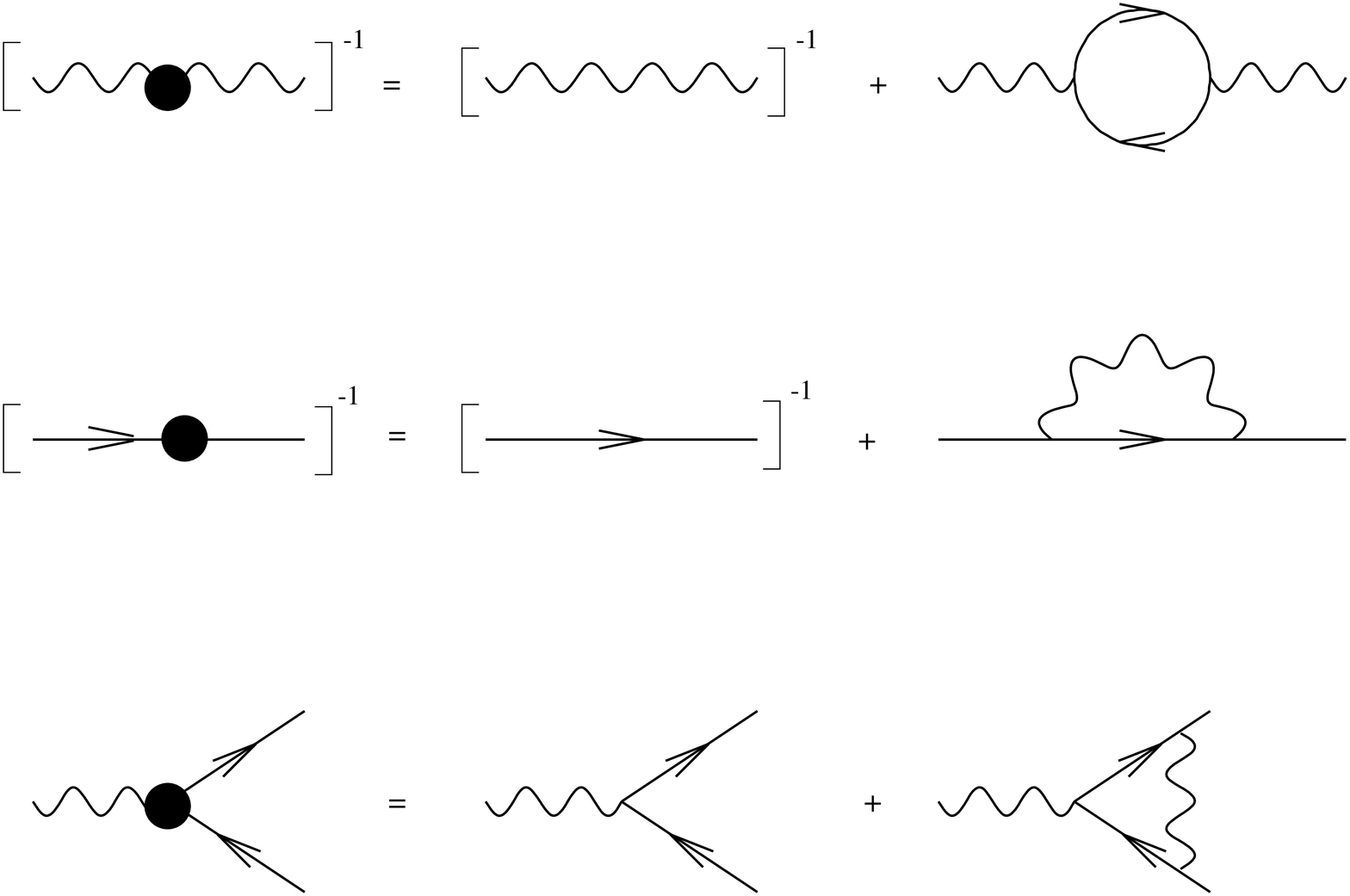}
\end{center}
\fcaption{Effective propagators and vertices for $QED$}
\label{effective}\end{figure}

The source of this paradox was argued at the time in particular
by Pisarski to be the breakdown of the loop expansion at high
temperature.\cite{r8} Further work
has resulted in what is now known as an effective expansion
in terms of hard thermal loops which aims to include in
a systematic manner all relevant loop effects to a given
order. In this expansion two scales of momenta are
relevant: hard ($\sim T$) and soft ($\sim gT$), with $g<<1$.
If an internal
momentum is hard then ordinary bare propagators and vertices
are sufficient, but if the momentum is soft then effective
propagators and vertices, as illustrated in Fig.~\ref{effective},
 must be used.
\par\noindent
The dominant contributions in the loop integrals come
from hard internal momenta, and hence these terms
are called hard thermal loops.\cite{klim,wel} These graphs enjoy some
remarkable properties such as gauge invariance
and so are interesting in their own right. They have
been studied from the point of
view of their relation to effective actions\cite{brat}${}^-$\cite{tay},
Chern--Simons theory\cite{nair}, and kinetic equations.\cite{ian,kelly}
\par\section{Successes of the effective expansion}
\begin{figure}[hb]
\begin{center}
\leavevmode
\epsfxsize=6 in
\epsfbox{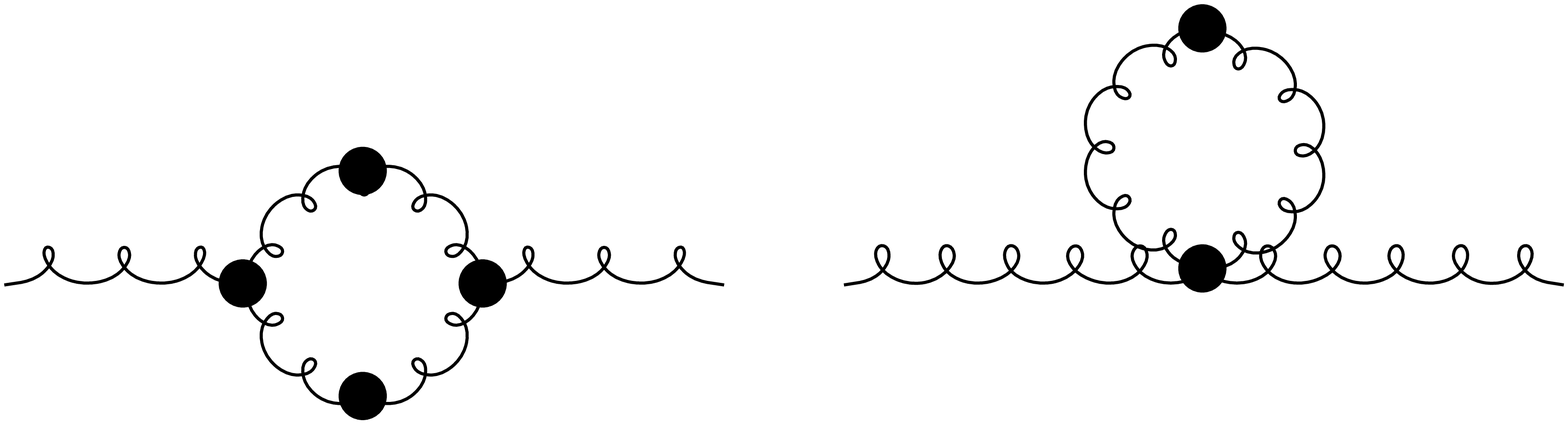}
\end{center}
\fcaption{Effective expansion of the gluon self--energy}
\label{gluon}\end{figure}

One of the first applications of this effective expansion
was to the calculation of damping rates for particles
at rest.\cite{gluon} For $QCD$ the relevant graphs are illustrated
in Fig.~\ref{gluon}, again with Faddeev--Popov ghost terms as required.
\par\noindent
The calculation of the damping rate from these graphs
is involved and requires numerical methods; nevertheless,
the result turns out to be positive (indicating
damped modes), gauge independent, and reasonable. Similar
calculations were also carried out for the fermion
damping rate.\cite{ferm,eric}
\par
A subtlety in these calculations was soon recognized.\cite{bks}
The gauge dependent piece of the damping rate, which
for covariant gauges is proportional to
\begin{equation}
  \zeta(k^2-m^2)^2\int\frac{dq}{q^4
    \left[(k+q)^2-m^2\right]},
\end{equation}
is finite as the mass--shell limit $k^2\to m^2$ is taken.
This would lead back to a gauge dependent damping rate.
A resolution to this paradox was later
proposed,\cite{eric,toni}${}^-$\cite{tang} where it
was shown that if one introduces an infrared cut--off
and then take the mass-shell limit before this cut--off
is removed, the gauge dependent contribution vanishes.
\par
Other calculations were subsequently done to test the
validity of this effective expansion. One such calculation
was the next--to--leading order correction to
the plasma frequency in the long--wavelength limit for $QCD$;
this involves the real parts of the graphs of
Fig.~\ref{gluon}, and is found to be\cite{schulz}
\begin{equation}
  \omega^2\sim \frac{1}{9}g^2NT^2\left[
    1-0.18g\sqrt{N}+\ldots\right].
\end{equation}
\par
\begin{figure}[ht]
\begin{center}
\leavevmode
\epsfxsize=3.2 in
\epsfbox{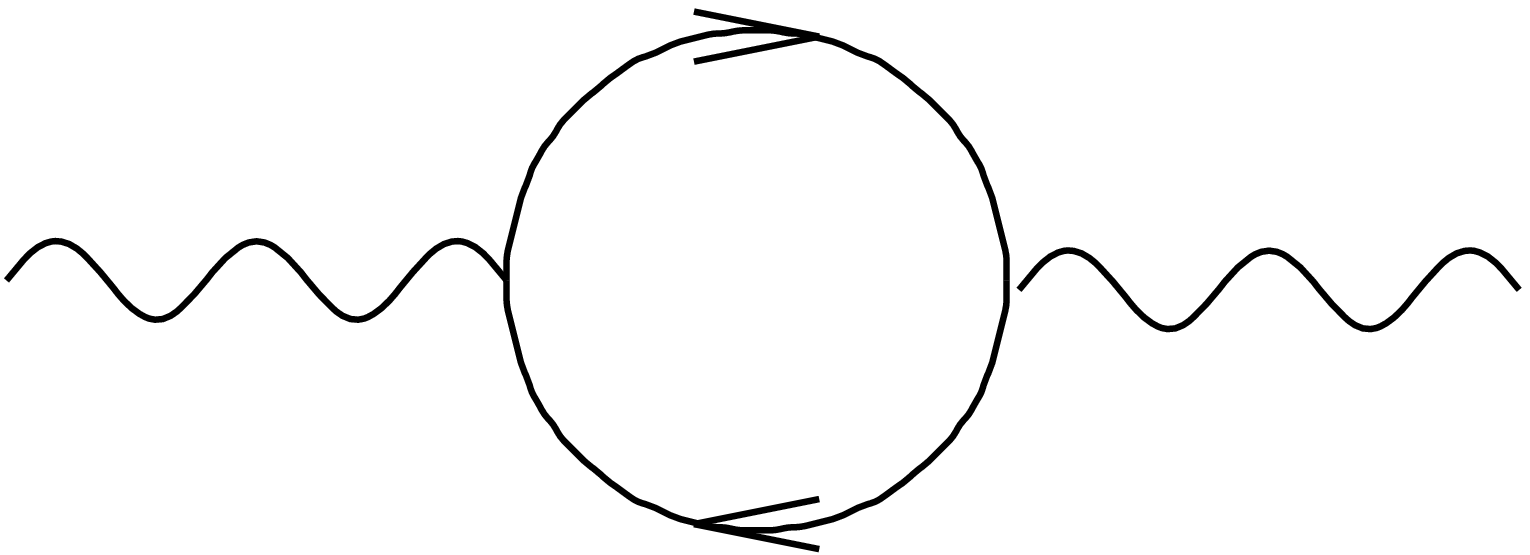}
\end{center}
\fcaption{Calculation of the photon production rate}
\label{photon}\end{figure}

Another calculation carried out was the photon production
rate for real hard photons
($k^2=0\,, k_\mu\sim T$),\cite{kap,nani}
which involves calculating
the imaginary part of the graph of Fig.~\ref{photon}.
\par\noindent
A general feature of calculations such as these is that internal
momenta can be hard or soft. When
hard bare propagators and vertices suffice, but when soft the
effective ones may be needed.
An intermediate scale ($\sim \sqrt{g}\,T$) is then introduced
to separate the contributions from each regime, and when
added together the dependence on this intermediate scale
must cancel. One finds for this calculation the result
\begin{equation}
  E\frac{dR}{d^3p}\sim\frac{\alpha^2T^2}{E}
  \,\ln\left(\frac{0.31E}{\alpha T}\right).
\end{equation}
Other successes of the effective expansion have also been
found.\cite{rolf}
\par\section{Problems with the effective expansion}
\begin{figure}[hb]
\begin{center}
\leavevmode
\epsfxsize=3.25 in
\epsfbox{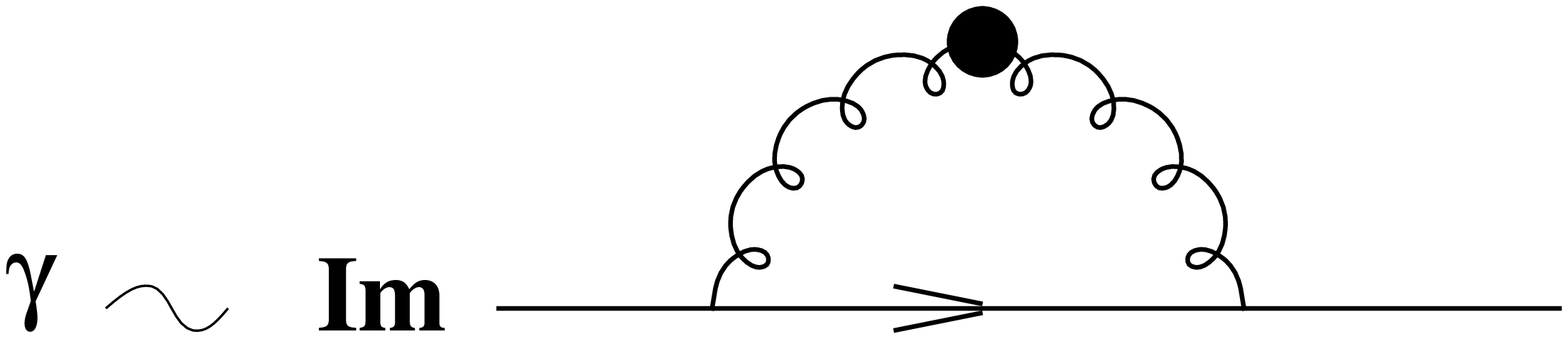}
\end{center}
\fcaption{Calculation of the fast fermion damping rate}
\label{fermion}\end{figure}

Although successful for the problems it addresses, the
effective expansion does not cure all the infrared problems
of hot perturbative gauge theories.
One example of this is the damping rate of a fast fermion
($E >> m >> T,\, v\to 1$), which involves the
calculation indicated in Fig.~\ref{fermion}.
\par\noindent
One finds for this that even with the effective propagator
an infrared divergence remains:
\begin{equation}
  \gamma\sim e^2T\int^{eT}\frac{dq}{q}.
\end{equation}
For $QCD$ this divergence can be removed by introduction of
a magnetic mass $m_{{\rm mag}}\sim g^2T$.\cite{cliff}${}^-$\cite{piz}
A resolution which also applies for $QED$, where
a magnetic mass is not expected to arise, was put forth by
Lebedev and Smilga.\cite{smilga}
They suggested to use an internal effective fermion
propagator of the form
\begin{equation}
  S(k)\sim\frac{i}{k_0-E+i\gamma},
\label{pole}\end{equation}
where $\gamma$ is the damping rate to be calculated.
The equation indicated in Fig.~\ref{fermion} with this
effective internal fermion propagator then
becomes an implicit equation for $\gamma$, which can be
solved to yield
\begin{equation}
  \gamma\sim e^2T\int^{eT}_\gamma\frac{dq}{q}\sim
  e^2T\ln\left(\frac{eT}{\gamma}\right)\sim
  e^2T\ln\left(\frac{1}{e}\right).
\end{equation}
\par\noindent
A subtlety with this calculation arose concerning the
proper on--shell condition to use; if the real
condition $k_0=E$ is used then the infrared divergence
disappears, but if the complex condition
$k_0=E-i\gamma$ is employed the divergence resurfaces.\cite{cjp}
A resolution to this difficulty could involve the use
of a more complicated form of the internal fermion
propagator with cuts rather than the simple pole of
Eq.(\ref{pole}).\cite{fast}
\par
Another calculation for which the effective
expansion is inadequate is the photon production
rate for soft real photons ($k^2=0\,, k_\mu\sim gT$).\cite{schiff,pat}
This involves again calculating
the imaginary part of the graph of Fig.~\ref{photon}.
With the appropriate effective propagators and vertices,
 however, a mass--shell
singularity remains.
\par
Further indications that the
effective expansion breaks down near the light--cone is
provided by calculating corrections to the
dispersion relation in
the vicinity of the light--cone. A relatively simple
illustration of this is scalar $QED$. There it is found that the
first--order corrections, which involve the
analogous graph to Fig.~\ref{photon}, diverge
logarithmically near the light--cone
($\sim \ln({\vec k}^{\,2}/k^2)$).\cite{kraemmer} As the ``tree--level''
effective propagator itself also diverges near the
light--cone ($\sim 1/\sqrt{k^2}$), this shows that
``loop'' corrections are no longer small in this regime,
and hence the effective expansion is breaking down.
\par
One final example of the inadequacy of the effective expansion
is the calculation of the next--to--leading order term of the
Debye mass. The definition of this mass involves
the potential\cite{joe}
\begin{equation}
  \Phi(r)\sim \int_{-\infty}^{+\infty}
  \frac{k\,dk}{k^2+\Pi_{00}(0,k)}
  \frac{\sin kr}{r},
\label{debye}\end{equation}
where $\Pi_{\mu\nu}$ is the gluon polarization tensor.
A gauge invariant definition of the Debye mass then follows
by considering the pole of Eq.(\ref{debye}):\cite{debye}
\begin{equation}
  m_D^2=\Pi_{00}(0, k^2=-m_D^2).
\end{equation}
At zeroth order one finds $m_D^2\sim g^2T^2$, but using
the effective expansion to calculate the next--to--leading
order term from the analogous graphs of Fig.~\ref{gluon}
leads to a divergent result unless a magnetic mass
$m_{{\rm mag}}\sim g^2T$ is included:\cite{debye,nieto}
\begin{equation}
  \frac{\delta m_D^2}{m_D^2}\sim
  g\ln\left(\frac{2m_D}{m_{\rm mag}}\right)\sim
  g\ln\left(\frac{1}{g}\right).
\end{equation}
Questions, though, have been raised over whether or
not a simple pole indicated in Eq.(\ref{debye}) is actually
present.\cite{peigne,kalash} Nevertheless,
as the magnetic mass for $QCD$ is expected to
be non--perturbative in nature, this provides another
example of the breakdown of the effective expansion.
\par\section{Beyond the hard thermal loop resummation}
Some attempts have been made to go beyond
the resummation techniques based on hard thermal loops
and address problems such as the ones discussed in the
last section. One fairly straightforward one which
enjoys some success concerns the behaviour of the effective
expansion near the light cone.\cite{kraemmer,flech}
 Specifically, one could imagine
including corrections to hard internal
lines, in the same manner such corrections are included
as in Fig.~\ref{effective} for soft lines.
Normally corrections to hard lines are of higher order than the corrections
to soft lines and thus can be ignored in lower--order
calculations. However, for processes near the light cone,
such as that of the dispersion relation of scalar $QED$
mentioned in the last section, corrections to hard lines turn out
to be as important as the soft line corrections, and indeed
remove the divergence found in the usual expansion using
just hard thermal loops. There has also been some work along
these lines to see if such corrections can remove the divergence
mentioned in the previous section for the soft photon
production rate.\cite{flech,nieg} Although these
corrections to hard
lines are known to improve the infrared behaviour in many
cases, a systematic approach based on their inclusion has not yet
been developed to the level of the usual hard thermal loop
expansion.
\par
\begin{figure}[hb]
\begin{center}
\leavevmode
\epsfxsize=2.2 in
\epsfbox{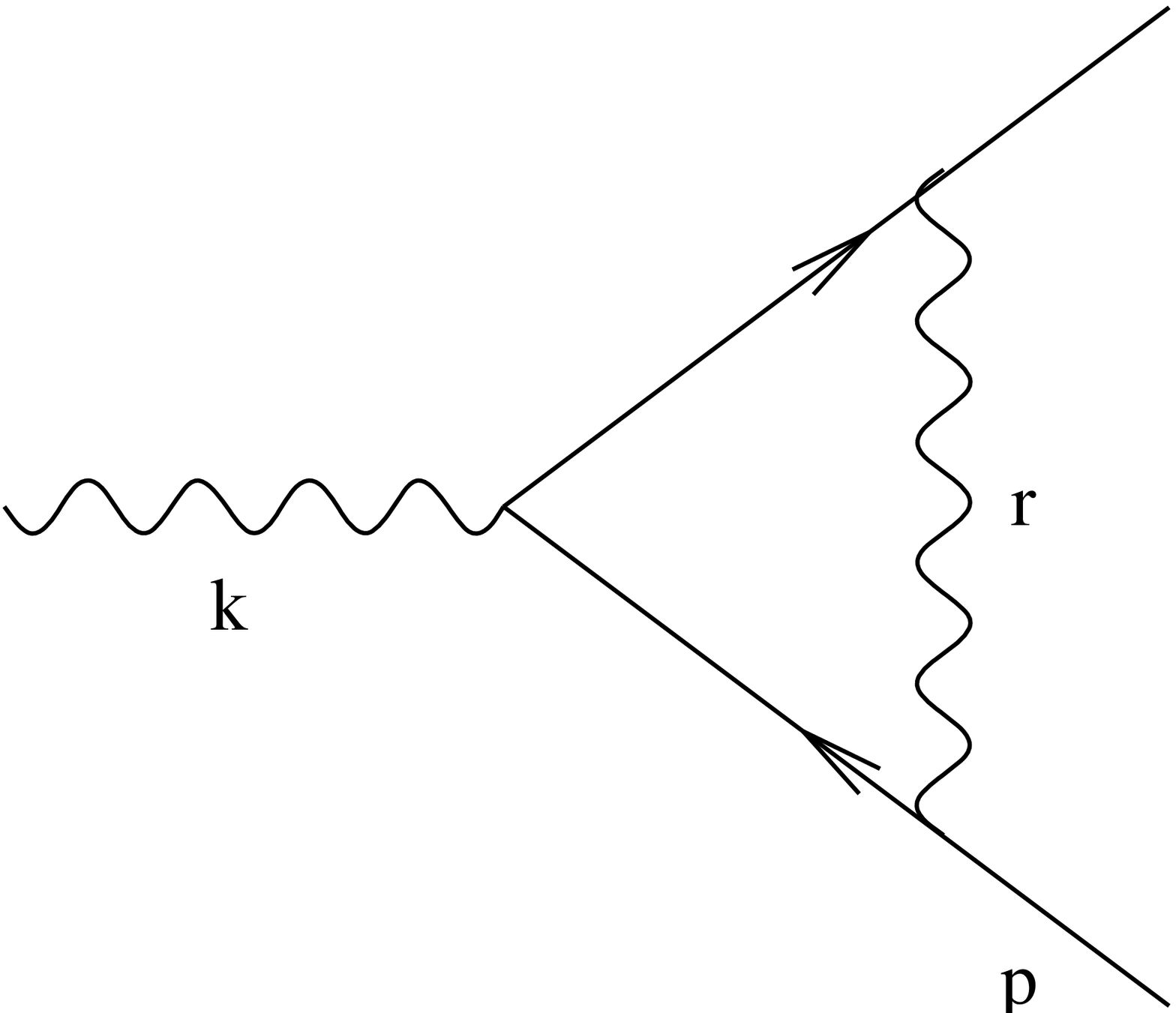}
\end{center}
\fcaption{One--loop vertex function}
\label{vertex}\end{figure}

Another approach which attempts to include the most infrared
singular terms perturbatively but whose relation
to the hard thermal loop expansion is not completely developed
is one based on an eikonal and/or ladder
approximation.\cite{cornwall,hou}
These general methods have been used with some success
in calculations
of the pressure,\cite{pressure} which as is well--known are plagued with
infrared divergences beyond certain orders, and it is hoped
they can be applied in other areas. Consider as
an illustration the one--loop vertex
function in $QED$ of Fig.~\ref{vertex}.
\par\noindent
In the limit that the internal photon momentum $r$ vanishes,
one finds that this function can be written in terms
of the one--loop self--energy $\Sigma(p)$ as
\begin{equation}
  \Gamma_\mu(k,p)= \frac{k_\mu+2p_\mu}{k^2+2k\cdot p}
  \left[\Sigma(p)-\Sigma(k+p)\right].
\label{solve}\end{equation}
This automatically satisfies the Ward identity
$k\cdot\Gamma(k,p)=\Sigma(p)-\Sigma(k+p)$, and so one
has in a sense ``solved'' the Ward identity for $\Gamma_\mu$
in terms of $\Sigma$ in this infrared limit. Although this relation
holds at the one--loop level, the functional relation
of Eq.(\ref{solve}) holds also at higher--loop orders. As well,
in a similar manner higher $n$--point functions can be
expressed in terms of the self--energy in this particular limit.
One could then attempt to use these approximations
in, for example, the Schwinger--Dyson equation for the
self--energy,
\begin{equation}
  \Sigma(p)=ie^2\int\,d^4k\,\gamma_\mu D^{\mu\nu}(p+k)
  S(k)\Gamma_\nu(p,k),
\label{sdeqn}\end{equation}
thereby obtaining a self--consistent equation for $\Sigma$.
Even though approximate, the momentum dependence makes
this equation difficult to solve, and one must do further
approximations in order to obtain a solution. One
approach along these lines is to use a relatively simple
form of a self--energy
function with some free parameters and then to employ
Eq.(\ref{sdeqn}) to determine these parameters.
One such approximation, based on the ansatz of
Eq.(\ref{pole}) involving a simple
pole structure of the propagator, has been tried.\cite{meg} However,
the assumption of a constant damping rate $\gamma$ has
proven to be too simplistic to lowest
non--trivial order.\cite{kraemmer}
More involved attempts thus must be made,
either by modifying the parameterized form of the
self--energy or else by using other
self--consistent approximations.\cite{kalash,peter}
\par
Another attempt to improve the usual loop expansion of
perturbation theory is the use of the renormalization
group equations.\cite{morley,mat}
These equations, which of course are
very successful at zero temperature, can be used to give
a running coupling constant $g(\tau,\kappa)$ which hopefully
will give information on the scaling of quantities with
temperature $\tau$ and momentum $\kappa$. This involves
calculating the finite temperature $\beta$--function from,
for example in the background field approach,\cite{bfmethod}
 the transverse
piece of the self--energy:
\begin{equation}
  \tau\frac{dg(\tau,\kappa)}{d\tau}=\beta_\tau=
  -\frac{g}{2\kappa^2}T\left.\frac{d\Pi_T(T,\kappa)}{dT}
    \right|_{T=\tau}.
\end{equation}
However, in high temperature $QCD$ this calculation is
plagued with difficulties at the one--loop level of
Fig.~\ref{oneloop}, among
which are problems with gauge and process
dependence.\cite{fuj}${}^-$\cite{van}
There are not even definite conclusions
on the sign of the $\beta$--function,
which would indicate asymptotic freedom and whether or not
the effective coupling constant decreases in strength as the
temperature rises. Attempts were made to improve on this
one--loop calculation by including the effects of the
hard thermal loops, as in Fig.~\ref{gluon}, as well
as inclusion by hand of a magnetic mass term for the transverse
gluons of order $g^2T$.\cite{per}
 These corrections give results in the
direction leaning towards asymptotic freedom but they are
inconclusive, as in particular
the problem of gauge dependence remain. Although one might
invoke arguments from the pinch
technique\cite{pinch1}${}^-$\cite{pinch3} or the
Vilkovisky--DeWitt effective
action\cite{vilk}${}^-$\cite{vilkdewitt}
 that certain gauges
might be ``preferred'', like the one--loop gluon damping
constant this gauge dependence probably indicates that the
expansion used is failing. As with the calculation of
the next--to--leading order term for the Debye mass described
in the previous section, definite results in this regard will
probably have to wait until a better understanding of scales
approaching the magnetic mass scale of order $g^2T$ is found,
if indeed this is possible in perturbation theory.
\par\section{Outlook}
Although successful for the problems it was designed for,
there are instances where the effective expansion based on
hard thermal loops breaks down due to the extreme infrared
behaviour. Examples found so far involve processes near
the light--cone and/or which begin to probe at the scale
$g^2T$ of the magnetic mass. Some of these problems may be
solved by inclusion of corrections to hard internal lines,
but this particular approach still needs to be developed
systematically to the same level as the hard thermal loop
expansion. Other problems, however, may require
expansions outside of the hard thermal loop expansion; such
work is still at a relatively exploratory level.
One might then
begin to wonder whether or not we have reached the calculational
limit of perturbation theory for hot gauge theories.
The view in favour of such a limit is supported by the
expected non--perturbative
nature of the magnetic mass of the gluon, and indeed work
has been developed recently in providing an interface
between perturbative expansions and non--perturbative
lattice gauge theory results.\cite{eff1,eff2} Although certainly of much
importance to high temperature $QCD$, problems in $QED$
would still remain, where a magnetic mass is not expected.
\par
One calculation that might help answer the question of
whether or not the perturbative limit has been reached
is based on the following argument.\cite{reynaud}
The ordinary tree--level
propagator becomes of the same order as the bare one--loop
self--energy of Fig.~\ref{oneloop} when the external
momentum is of order $gT$. This necessitates the use of the
effective expansion as in Fig.~\ref{gluon}, which in particular
involves the effective propagator of Fig.~\ref{effective}.
One might then ask when this effective
propagator becomes of the same order as the effective
one--loop terms of Fig.~\ref{gluon}. Preliminary
results indicate that this might happen at a scale
of order $g^{4/3}T$, which is sooner than the
probably non--perturbative $g^2T$ scale.
Although more work is needed to
complete this calculation, if these results are
correct it could mean there is still room for perturbation
theory applied to hot gauge fields.
\clearpage
\par\section{Acknowledgments}
We were all saddened recently to lose two
very active members of the thermal field
theory community -- Tanguy Altherr and Hiroomi Umezawa.
As well as being a personal loss to family and friends,
their contributions to physics and to workshops
such as these will be greatly missed.
\par
I would like to thank the organizers of this workshop,
especially Profs.~Gui and Khanna, for the opportunity to
attend, both in terms of physics and having the chance
to visit such an interesting country. Discussions at this
workshop and elsewhere on this topic were very beneficial.
This work was supported by the Natural Sciences and Engineering
Research Council of Canada and by a travel grant from
the University of Winnipeg.
\par
\end{document}